\documentclass[usenatbib]{mnras}

\usepackage[T1]{fontenc}
\usepackage{ae,aecompl}
\usepackage{graphicx}
\usepackage{amsmath,amssymb}
\usepackage{color,ulem}
\usepackage[compatibility=false]{caption }
\usepackage{natbib}

\usepackage{verbatim}

\definecolor{webgreen}{rgb}{0,.5,0}
\definecolor{webbrown}{rgb}{.6,0,0}

\begin{document}
\title[SMBHs, hot atmospheres, and total masses of ETGs]{Correlations between supermassive black holes, hot atmospheres, and the total masses of early type galaxies}
\author[]{K. Lakhchaura$^{1,2}$\thanks{lakhchaura.k@gmail.com}, N. Truong$^{1}$, N. Werner$^{1,3,4}$\\
$^1$MTA-E\"otv\"os University Lend\"ulet Hot Universe Research Group, P\'azm\'any P\'eter s\'et\'any 1/A, Budapest, 1117, Hungary \\
$^2$MTA-ELTE Astrophysics Research Group, P\'azm\'any P\'eter s\'et\'any 1/A, Budapest, 1117, Hungary \\
$^3$Department of Theoretical Physics and Astrophysics, Faculty of Science, Masaryk University, Kotl\'a\v{r}sk\'a 2, Brno, 611 37, Czech Republic \\
$^4$School of Science, Hiroshima University, 1-3-1 Kagamiyama, Higashi-Hiroshima 739-8526, Japan \\
}
\date{Accepted 2019 July 9. Received 2019 June 17; in original form 2019 April 23.}
\maketitle

\begin{abstract} 
We present a study of relations between the masses of the central supermassive black holes (SMBHs) and the atmospheric gas temperatures and luminosities 
measured within a range of radii between $R_{\rm e}$ and 5$R_{\rm e}$, for a sample of 47 early-type galaxies observed by the {\it Chandra X-ray Observatory}. 
We report the discovery of a tight correlation between the atmospheric temperatures of the brightest cluster/group galaxies (BCGs) and their central SMBH masses. Furthermore, our hydrostatic analysis reveals an approximately linear correlation 
between the total masses of BCGs ($M_{\rm tot}$) and their central SMBH masses ($M_{\rm BH}$). 
State-of-the-art cosmological simulations show that the SMBH mass could be determined by the binding energy of the halo through radiative feedback during the rapid black hole growth by accretion, while for the most massive galaxies mergers are the chief channel of growth.
In the scenario of a simultaneous growth of central SMBHs and their host galaxies through mergers, the observed linear correlation could be a natural consequence of the central limit theorem. 
\end{abstract}

\begin{keywords}
galaxies: evolution -- galaxies: formation -- galaxies: active -- X-rays: galaxies
\end{keywords}

\section{Introduction} 
\label{sec:intro}

The masses of central supermassive black holes (SMBHs) are known to correlate with the luminosities and velocity dispersions 
of the stellar bulge components of their host galaxies, suggesting a coevolution of the two \citep[][]{Magorrian1998,Gebhardt2000,FerrareseMerritt2000,KormendyHo2013,Saglia2016}.
Most of the previous studies have explained the observed correlations based on the active galactic nucleus (AGN) feedback induced by the gas accreted 
onto the SMBH \citep{SilkRees1998,King2003,Churazov2005}. This picture relies on the assumption that the AGN feedback couples efficiently to the circum-SMBH gas, regulating the growth of 
both the SMBH and the bulge component.

However, the origin of the observed correlations may also be non-causal \citep[][]{Peng2007,JahnkeMaccio2011,Volonteri2011}. This view is based on the central limit theorem, 
which  suggests that a sufficient number of  mergers would 
lead to the growth of both the mass of the SMBH ($M_{\rm BH}$) and the host galaxy's total mass ($M_{\rm tot}$), resulting in a linear correlation between the two. As a consequence, 
any quantity that traces $M_{\rm tot}$, must be a good proxy for $M_{\rm BH}$. 

Previous studies \citep[see][]{Ferrarese2002,Baes2003} investigated the connection between $M_{\rm BH}$ and $M_{\rm tot}$ (which also includes the mass of the underlying dark matter halo), 
based on a correlation between the circular velocity ($\nu_{\rm c}$; related to $M_{\rm tot}$) and the stellar velocity dispersion ($\sigma_{\rm e}$; related to $M_{\rm BH}$). 
This  was later rejected based on the argument that it is a result of a `conspiracy' between baryons and dark matter that also makes the rotation curves flat \citep{KormendyBender2011}. 
Recently, a relationship was also suggested between the $M_{\rm BH}$ of the brightest cluster/group galaxies (BCGs) and the virial masses of their host clusters/groups, estimated from the temperature of the intra-cluster medium, albeit with a relatively large scatter of $\sim$0.4-0.9 dex \citep[see][]{Bogdan2018,Phipps2019}.

The growth of SMBHs has also been investigated using cosmological simulations. Using the EAGLE simulations, \citet{McAlpine2018} 
\citep[see also][Fig. 2 \& 6]{Bower2017} identify three distinct phases of a SMBH's growth depending on the mass of the host galaxy ($M_{\rm gal}$): i) in low-mass galaxies ($M_{\rm gal}<<10^{12}$ $M_{\odot}$) SMBH's growth is 
hindered by supernovae driven winds, ii) at $M_{\rm gal}\sim 10^{12}$ $M_{\odot}$, SMBHs undergo a fast growth phase via gas accretion, thanks to the formation of hot atmospheres that efficiently prevent the 
supernovae-driven winds from buoyantly rising out of the galaxy's central region, and iii) in the largest galaxies ($M_{\rm gal}\gtrsim 10^{12}$ $M_{\odot}$), SMBHs are massive enough to inject sufficient energy via 
kinetic mode so that they can regulate the inflow of gas and then retire to a Bondi-like low accretion rate. \citet{Weinberger2018}, using the Illustris TNG300 simulations, find a similar pattern of SMBH growth, 
adding that for the most massive SMBHs ($M_{BH}>10^{8.5}$ $M_{\odot}$) mergers become the chief channel of growth \citep[see also][]{Bassini2019}.

In this work, we have investigated the correlation between $M_{\rm BH}$ and the hot galactic atmospheres of the host galaxies, based on the analysis of their \textit{Chandra} X-ray observations\footnote{An independent paper submitted for publication by \citet{Gaspari2019} investigates the relation between the masses of supermassive
black holes and hot atmospheres based on literature measurements and
proposes a different interpretation.}. The details of the sample are given in \S\ref{sec:sampl_sel}; the data reduction and analysis details are given in \S\ref{sec:xray_data_red_analys}; the results are described in \S\ref{sec:results} and discussed in \S\ref{sec:discussion}.

\section{Sample and Data}
\subsection{Sample Selection}
\label{sec:sampl_sel}

We have analysed archival \textit{Chandra} X-ray observations for a sample of early-type galaxies, selected based on the availability of $M_{\rm BH}$ estimates, obtained using  
stellar/gas dynamics (\citealp[from][]{KormendyHo2013} \citealp[,][]{Saglia2016}  \citealp[and][]{vandenBosch2016} ), and archival \textit{Chandra} X-ray observations. 
The black hole masses for galaxies undergoing mergers can be severely underestimated \citep[see][]{KormendyHo2013}. Therefore, we excluded the galaxies 
with recent/ongoing mergers (NGC 1316, NGC 2960, NGC 4382, NGC 5128, IC 1481, NGC 741, NGC 3379). 
As done in \citet{KormendyHo2013}, we also excluded the galaxies with SMBH masses estimated based on kinematics of ionized gas but without the broad emission 
line widths being taken into account (NGC 2778, NGC 3607, NGC 4261, NGC 4459, NGC 6251, NGC 7052, A1836 BCG, Cygnus A and NGC 4596) as that can also lead to 
an underestimation of the SMBH mass. We also excluded the well known shell galaxies NGC 3923 and IC 1459. The emission in NGC 821 and NGC 3115 was found to be heavily contaminated 
by the emission from unresolved point sources and therefore were excluded from the sample. The X-ray spectra extracted for NGC 5845, NGC 4026, NGC 4564 and NGC 7457 
had very few counts, and therefore were deemed unusable for the analysis. The SMBH masses for 3C 449 and NGC 3945 were found to be exceedingly uncertain and therefore were 
excluded from the sample. The properties of the finally selected 47 galaxies, including 18 brightest central 
cluster/group galaxies (BCGs), 16 non-BCG elliptical galaxies, and 13 lenticular galaxies, are listed in Table \ref{tab:properties}.
\newline
\begin{table*}
 \caption{The names, distances and effective radii \citep[$R_{\rm e}$, from][]{vandenBosch2016}, hot gas temperatures ($kT$; within $R_{\rm e}$), 0.5--7.0 keV X-ray luminosities ($L_{\rm X}$; within $R_{\rm e}$), black hole masses ($M_{\rm BH}$), effective stellar velocity dispersions ($\sigma_{\rm e}$), 
 hydrostatic mass estimates ($M_{\rm HE}$; within 5$R_{\rm e}$) and references for $M_{\rm BH}$/$\sigma_{\rm e}$, 
 for the 18 BCGs, 16 non-BCG ellipticals, 13 lenticulars and 2 massive compact relic galaxies, studied in this paper. 
 }
\label{tab:properties}
\centering
{\scriptsize
\setlength\tabcolsep{4pt} 
\begin{tabular}{c c c c c c c c c}
\hline
Name & Distance (Mpc) & $R_{\rm e}$ (kpc) & $kT$ (keV) & $L_{\rm X}$ (10$^{42}$ erg s$^{-1}$) & $M_{\rm BH}$ (10$^9$ $M_{\odot}$) & $\sigma_{\rm e}$ (km s$^{-1}$) & $M_{\rm HE}$ (5$R_{\rm e}$) (10$^{12}$ $M_{\odot}$) &  $M_{\rm BH}$/$\sigma_{\rm e}$ Ref.\\
\hline
\multicolumn{8}{c}{BCGs}\\
\hline
NGC 315  & 57.7 & 10.2 & 0.70$\pm$0.01 & 0.120$\pm$0.035 & $0.832\pm0.594$ & 309$\pm$29 & 1.83$\pm$0.39 &  \citet{vandenBosch2016}\\
NGC 1052 & 18.1 & 2.2 & 0.32$\pm$0.02 & 0.0022$\pm$0.0002 & $0.174\pm0.116$ & 191$\pm$4 & 0.09$\pm$0.02 & \citet{vandenBosch2016}\\
NGC 1332 & 22.3 & 4.8 & 0.62$\pm$0.02 & 0.008$\pm$0.004 & $1.47^{+0.21}_{-0.20}$ & 328$\pm$9 & 0.75$\pm$0.08 & \citet{KormendyHo2013}\\
NGC 1399 & 20.9 & 5.5 & 1.013$\pm$0.004 & 0.138$\pm$0.005 & $0.88^{+0.90}_{-0.45}$ & 315$\pm$3 & 1.49$\pm$0.03 & \citet{KormendyHo2013}\\
NGC 1407 & 28.0 & 9.3 & 0.89$\pm$0.01 & 0.093$\pm$0.006 & $4.65^{+0.73}_{-0.41}$ & 276$\pm$2 & 1.83$\pm$0.06 & \citet{KormendyHo2013}\\
NGC 1550 & 51.6 & 4.6 & 1.15$\pm$0.02 & 0.551$\pm$0.039 & $3.87^{+0.61}_{-0.71}$ & 270$\pm$10 & 0.91$\pm$0.01 & \citet{KormendyHo2013}\\
NGC 3091 & 51.2 & 7.8 & 0.83$\pm$0.01 & 0.165$\pm$0.023 & $3.72^{+0.11}_{-0.51}$ & 297$\pm$12 & 0.94$\pm$0.10 & \citet{KormendyHo2013}\\
NGC 3842 & 92.2 & 12.9 & 1.06$\pm$0.03 & 0.082$\pm$0.006 & 9.10$\pm$2.91 & 270$\pm$27 & 5.15$\pm$0.84 & \citet{Saglia2016}\\
NGC 4291 & 26.6 & 2.8 & 0.49$\pm$0.03 & 0.015$\pm$0.006 & 0.98$\pm$0.31 & 242$\pm$12 & 0.20$\pm$0.02 & \citet{KormendyHo2013}\\
NGC 4486 & 16.7 & 6.6 & 1.655$\pm$0.001 & 1.444$\pm$0.004 & $6.15^{+0.38}_{-0.37}$ & 324$\pm$20 & 1.705$\pm$0.002 & \citet{KormendyHo2013}\\
NGC 4697 & 12.5 & 4.4 & 0.33$\pm$0.01 & 0.0015$\pm$0.0004 & 0.20$\pm$0.05 & 177$\pm$8 & 0.21$\pm$0.08 & \citet{KormendyHo2013}\\
NGC 5077 & 38.7 & 3.3 & 0.60$\pm$0.05 & 0.009$\pm$0.001 & $0.86^{+0.44}_{-0.45}$ & 222$\pm$11 & 0.63$\pm$0.19 & \citet{KormendyHo2013}\\
NGC 5328 & 64.1 & 8.7 & 0.91$\pm$0.06 & 0.168$\pm$0.020 & $4.677\pm1.723$ & 331$\pm$15 & 1.55$\pm$0.19 & \citet{vandenBosch2016}\\
NGC 5419 & 56.2 & 18.2 & 1.29$\pm$0.03 & 0.315$\pm$0.019 & 7.24$\pm$2.40 & 367$\pm$9 & 9.00$\pm$0.90 & \citet{Saglia2016}\\
NGC 5490 & 65.2 & 5.0 & 0.98$\pm$0.05 & 0.084$\pm$0.010 & $0.537\pm0.433$ & 257$\pm$24 & 1.16$\pm$0.19 & \citet{vandenBosch2016}\\
NGC 5813 & 32.2 & 9.1 & 0.689$\pm$0.001 & 0.496$\pm$0.005 & 0.71$\pm$0.09 & 230$\pm$11 & 1.01$\pm$0.01 & \citet{Saglia2016}\\
NGC 5846 & 24.9 & 6.3 & 0.715$\pm$0.003 & 0.172$\pm$0.007 & 1.10$\pm$0.15 & 237$\pm$12 & 0.90$\pm$0.01 & \citet{Saglia2016}\\
NGC 7619 & 51.5 & 7.4 & 0.87$\pm$0.01 & 0.124$\pm$0.001 & $2.30^{+0.15}_{-0.11}$ & 292$\pm$5 & 1.51$\pm$0.05 & \citet{KormendyHo2013}\\
\hline
\multicolumn{8}{c}{Non-BCGs}\\
\hline
NGC 541 	& 63.7 & 8.3 & 0.72$\pm$0.03 & 0.027$\pm$0.009 & 0.389$\pm$0.305 & 191$\pm$4 & -- & \citet{vandenBosch2016}\\
NGC 1374 & 19.2 & 1.7 & 0.86$\pm$0.19 & 0.0004$\pm$0.0002 & $0.59^{+0.06}_{-0.05}$ & 167$\pm$3 & -- & \citet{KormendyHo2013}\\
NGC 2892 & 86.2 & 6.8 & 0.64$\pm$0.09 & 0.072$\pm$0.014 & 0.269$\pm$0.068 & 295$\pm$20 & -- & \citet{vandenBosch2016}\\
NGC 3377 & 11.0 & 2.3 & 0.19$\pm$0.03 & 0.0001$\pm$0.00003 & 0.18$\pm$0.09 & 145$\pm$7 & -- & \citet{KormendyHo2013}\\
NGC 3608 & 22.8 & 3.0 & 0.33$\pm$0.06 & 0.0013$\pm$0.0003 & 0.47$\pm$0.10 & 182$\pm$9 & -- & \citet{KormendyHo2013}\\
NGC 3862 & 84.6 & 12.0 & 0.38$\pm$0.04 & 0.037$\pm$0.005 & 0.257$\pm$0.219 & 209$\pm$14 & -- & \citet{vandenBosch2016}\\
NGC 4374 & 18.5 & 3.7 & 0.73$\pm$0.01 & 0.043$\pm$0.003 & $0.93^{+0.10}_{-0.09}$ & 296$\pm$14 & -- & \citet{KormendyHo2013}\\
NGC 4472 & 17.1 & 7.8 & 0.987$\pm$0.001 & 0.311$\pm$0.003 & $2.54^{+0.58}_{-0.10}$ & 300$\pm$7 & -- & \citet{KormendyHo2013}\\
NGC 4473 & 15.2 & 2.4 & 0.38$\pm$0.07 & 0.0005$\pm$0.0001 & 0.09$\pm$0.05 & 190$\pm$9 & -- & \citet{KormendyHo2013}\\
NGC 4552 & 15.3 & 2.8 & 0.65$\pm$0.01 & 0.0200$\pm$0.0002 & 0.50$\pm$0.06 & 252$\pm$12 & -- & \citet{Saglia2016}\\
NGC 4621 & 18.3 & 3.4 & 0.23$\pm$0.06 & 0.0003$\pm$0.0001 & 0.40$\pm$0.08 & 225$\pm$11 & -- & \citet{Saglia2016}\\
NGC 4649 & 16.5 & 7.9 & 0.925$\pm$0.002 & 0.098$\pm$0.002 & $4.72^{+1.04}_{-1.05}$ & 380$\pm$19 & -- & \citet{KormendyHo2013}\\
NGC 4751 & 26.9 & 3.3 & 0.40$\pm$0.09 & 0.011$\pm$0.002 & $2.44^{+0.12}_{-0.37}$ & 355$\pm$14 & -- & \citet{KormendyHo2013}\\
NGC 5576 & 25.7 & 4.9 & 0.31$\pm$0.08 & 0.0009$\pm$0.0003 & $0.27^{+0.07}_{-0.08}$ & 183$\pm$9 & -- & \citet{KormendyHo2013}\\
NGC 7626 & 38.1  & 7.4 & 0.71$\pm$0.03  & 0.039$\pm$0.013 & 0.380$\pm$0.289 & 234$\pm$11 & -- & \citet{vandenBosch2016}\\
UGC 1841 & 74.9  & 30.9 & 1.31$\pm$0.03   & 0.418$\pm$0.058 & 0.295$\pm$0.129 & 295$\pm$27 & -- & \citet{vandenBosch2016}\\
\hline
\multicolumn{8}{c}{Lenticulars}\\
\hline
NGC 193 & 49.7 & 6.8 & 0.77$\pm$0.03 & 0.010$\pm$0.001 & 0.251$\pm$0.185 & 187$\pm$17 & -- & \citet{vandenBosch2016}\\
NGC 383 & 59.2 & 12.3 & 0.93$\pm$0.02 & 0.060$\pm$0.013 & 0.575$\pm$0.424 & 240$\pm$17 & -- & \citet{vandenBosch2016}\\
NGC 524 & 24.2 & 3.6 & 0.50$\pm$0.05 & 0.011$\pm$0.003 & $0.87^{+0.09}_{-0.05}$ & 247$\pm$12 & -- & \citet{KormendyHo2013}\\
NGC 1023 & 10.8 & 3.1 & 0.32$\pm$0.01 & 0.00056$\pm$0.00004 & 0.042$\pm$0.004 & 205$\pm$10 & -- & \citet{KormendyHo2013}\\
NGC 3245 & 21.4 & 2.5 & 0.33$\pm$0.05 & 0.0018$\pm$0.0004 & $0.24^{+0.03}_{-0.08}$ & 205$\pm$10 & -- & \citet{KormendyHo2013}\\
NGC 3585 & 20.5 & 6.3 & 0.31$\pm$0.02 & 0.0023$\pm$0.0002 & $0.33^{+0.15}_{-0.06}$ & 213$\pm$11 & -- & \citet{KormendyHo2013}\\
NGC 3665 & 34.7 & 6.3 & 0.40$\pm$0.07 & 0.011$\pm$0.002 & 0.575$\pm$0.118 & 219$\pm$10 & -- & \citet{vandenBosch2016}\\
NGC 4036 & 19.0 & 2.7 & 0.33$\pm$0.05 & 0.0019$\pm$0.0004 & 0.077$\pm$0.064 & 182$\pm$8 & -- & \citet{vandenBosch2016}\\
NGC 4203 & 14.1 & 2.6 & 0.29$\pm$0.04 & 0.0011$\pm$0.0002 & 0.066$\pm$0.040 & 129$\pm$6 & -- & \citet{vandenBosch2016}\\
NGC 4342 & 22.9 & 0.6 & 0.63$\pm$0.10 & 0.0003$\pm$0.0001 & $0.45^{+0.27}_{-0.15}$ & 225$\pm$11 & -- & \citet{KormendyHo2013}\\
NGC 4477 & 20.8 & 4.9 & 0.38$\pm$0.01 & 0.0107$\pm$0.0004 & 0.035$\pm$0.024 & 148$\pm$7 & -- & \citet{vandenBosch2016}\\
NGC 4526 & 16.4 & 3.5 & 0.34$\pm$0.02 & 0.0022$\pm$0.0002 & $0.45^{+0.14}_{-0.10}$ & 222$\pm$11 & -- & \citet{KormendyHo2013}\\
NGC 6861 & 27.3 & 2.1 & 0.80$\pm$0.02 & 0.018$\pm$0.003 & $2.10^{+0.63}_{-0.10}$ & 389$\pm$3 & -- & \citet{KormendyHo2013}\\
\hline
\multicolumn{8}{c}{Compact Relics}\\
\hline
Mrk 1216 & 97.0 & 2.3 & 0.91$\pm$0.03 & 0.28$\pm$0.09 & 4.9$\pm$1.7 & 324.1$\pm$15.6 & -- & \citet{Walsh2017}\\
NGC 1277 & 56.9 & 1.4 & 1.02$\pm$0.06 & 0.20$\pm$0.01 & 4.9$\pm$1.6 & 250.0$\pm$15.3 & -- & \citet{Walsh2016}\\
\hline
\end{tabular}
\footnotetext[2]{$M_{\rm BH}$ reference 1: \citet{KormendyHo2013}, 2: \citet{Saglia2016}, 3: \citet{Walsh2016}, 
 4: \citet{Walsh2017} }}
\end{table*}

\subsection{X-ray Data Reduction and Analysis}
\label{sec:xray_data_red_analys}


The \textit{Chandra} X-ray observations of all the galaxies were obtained from the \textit{High Energy Astrophysics Science Archive Research Centre (HEASARC)}. 
We used the \textit{Chandra Interactive Analysis of Observations (CIAO)} software version 4.9 \citep{Fruscione2006} and CALDB version 4.7.3 for all the data reduction and spectral extraction. For the spectral analysis, we used the X-ray spectral fitting package \textit{XSPEC} version 12.9.1 \citep{Arnaud1996} with ATOMDB version 3.0.7 \citep{foster2012}. The data were reprocessed using the \textit{CIAO} tool \textit{chandra\_repro} and time periods
with strong background flares were removed using the \textit{CIAO} script \textit{lc\_clean}.
Point sources in the field were detected using the \textit{CIAO} task \textit{wavdetect}, verified by
visual inspection of the X-ray images and finally removed from all the event files.


The X-ray spectra were extracted from circular regions within the half-light radius ($R_{\rm e}$) i.e. the radius at 
which the integrated light is half of the total light emitted. For all the galaxies, the $R_{\rm e}$ values were taken from \citet{vandenBosch2016}. 
We also extracted spectra for the central 10 kpc, 3$R_{\rm e}$ and 5$R_{\rm e}$ regions. For some of the galaxies, the 3$R_{\rm e}$ or 5$R_{\rm e}$ regions 
were found to be outside the field of view of the detector. 
These spectra were not used in the analysis (e.g., NGC 524). For NGC 3842, the deprojected temperature profile shows that the emission in the 3$R_{\rm e}$ or 5$R_{\rm e}$ regions are mostly 
dominated by the high temperature cluster plasma and therefore we did not use its spectra from these regions.


For all of the galaxies (except NGC 3842), the standard \textit{Chandra} blanksky files were used to extract the background spectra. The matching blanksky event files were selected and 
reprojected to match the source observation coordinates. The time-dependent particle background levels were matched by scaling the blanksky spectra by the 9.5--12 keV count rate ratios of the source and blanksky observations. 
Due to projection effects, for NGC 3842 (A1367 BCG), the galactic X-ray emission was highly 
contaminated by the emission from the surrounding high temperature ($\sim$4 keV) intracluster medium. Therefore, for NGC 3842, we used the data 
extracted from an annular region, just outside $R_{\rm e}$, for the background spectrum.


The X-ray spectrum for each galaxy was modeled with an absorbed single-temperature \textit{apec} component that describes a thermal plasma 
in collisional ionization equilibrium. The temperature, metallicity, 
and spectrum normalization were included as free parameters \citep[for details, see][]{Lakhchaura2018}. For the neutral hydrogen column density, we used the 
\textit{Swift} Galactic $N_{\rm H}$ tool, which uses the method of \cite{Willingale2013}. The redshift was taken from the SIMBAD database \citep{SIMBAD}. An additional 
thermal bremsstrahlung component with a fixed temperature of 7.3 keV was used to model 
the unresolved point sources in the field \citep{Irwin2003}. The X-ray temperatures and the 0.5--7.0 keV luminosities for all 47 galaxies in our sample, determined within $R_{\rm e}$, are given in Table \ref{tab:properties}.

\section{Results}
\label{sec:results}

\begin{figure*}
\centering
  \includegraphics[width=0.7\linewidth]{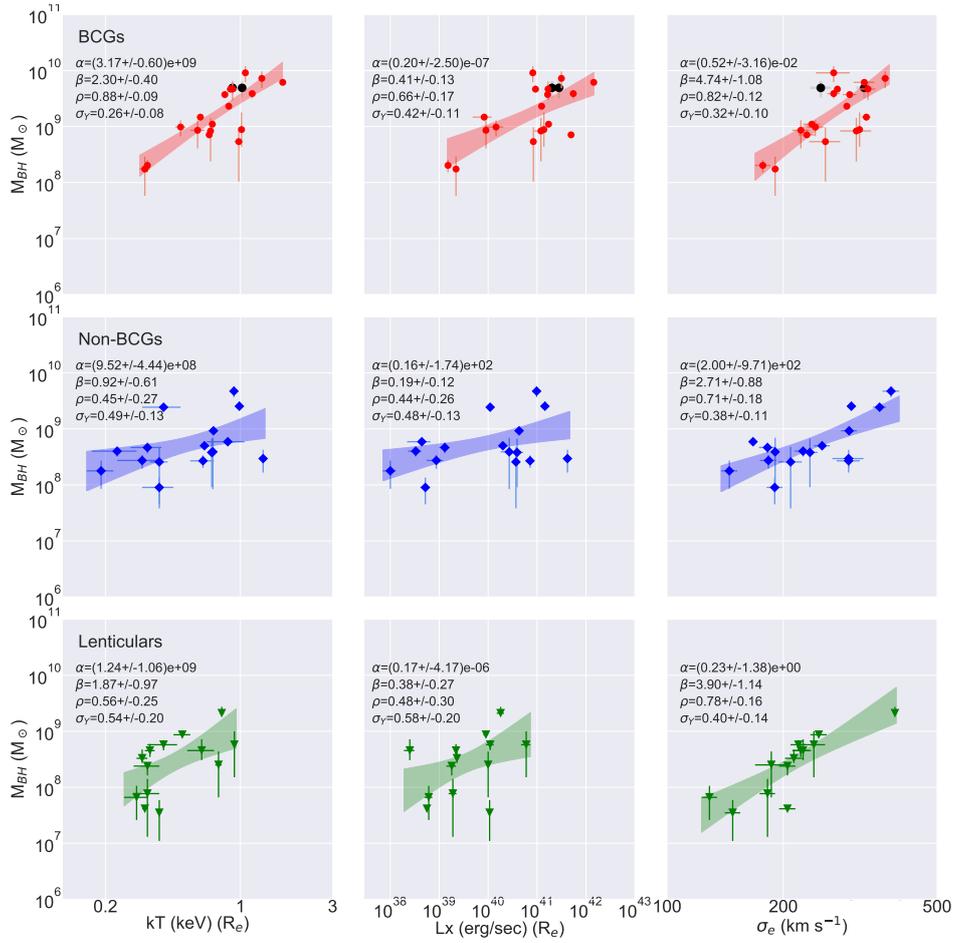}
\caption{Correlations of the SMBH mass ($M_{\rm BH}$) with the temperature ($kT$; left) and 0.5--7.0 keV luminosity ($L_{\rm X}$; centre) of the hot gas from within the half light radius ($R_{\rm e}$), and the stellar velocity dispersion 
($\sigma_{\rm e}$; right) for the BCGs (top row), non-BCG ellipticals (middle row) and lenticulars (bottom row). The intercept ($\alpha$), slope ($\beta$), correlation coefficient ($\rho$) and intrinsic scatter 
($\sigma_{\rm Y}$; in dex units), obtained from the log-log correlation analyses ($Y=\alpha X^{\beta}$), and their 68\% uncertainties are given in the insets. The shaded areas show the 68\% confidence regions for the correlations. 
The black circles in the top row show the massive compact relic galaxies, Mrk 1216 and NGC 1277.}
\label{fig:MBH_Re_ correlns_full_sample}
\end{figure*}

The correlations between $M_{\rm BH}$, atmospheric gas temperature ($kT$), and X-ray luminosity ($L_{\rm X}$) of the galactic atmospheres were determined 
using the  \textit{linmix} package \citep{Kelly2007} in \textit{Python} that uses a Bayesian linear regression method that accounts for measurement errors in both axes.
The results of the correlation analyses, for the BCGs, non-BCGs, 
and lenticular galaxies within $R_{\rm e}$, are shown in Fig. \ref{fig:MBH_Re_ correlns_full_sample}. For comparison, the figure also shows the 
correlation of $M_{\rm BH}$ with the effective stellar velocity dispersion (\citealp[$\sigma_{\rm e}$; taken from][]{KormendyHo2013} \citealp[,][]{Saglia2016}  \citealp[and][]{vandenBosch2016}). 
The results of similar analyses within the central 10 kpc, 3$R_{\rm e}$ and 5$R_{\rm e}$ regions are shown in Figs. \ref{fig:MBH_10kpc_ correlns_full_sample}, 
\ref{fig:MBH_3Re_ correlns_full_sample} and \ref{fig:MBH_5Re_ correlns_full_sample}, respectively.

From Fig. \ref{fig:MBH_Re_ correlns_full_sample}, we see that for the BCGs, $M_{\rm BH}$ correlates strongly with $kT$ within $R_{\rm e}$ 
($M_{\rm BH}$/M$_{\odot}=$(3.2$\pm$0.6)$\times 10^9$ ($kT$/keV)$^{2.3\pm0.4}$) with a correlation coefficient of 0.88$\pm$0.08 and intrinsic scatter of 0.26$\pm$0.08 dex. 
For the non-BCG elliptical and lenticular galaxies, the correlation is found to be weaker. $M_{\rm BH}$ correlates moderately with the X-ray 
luminosity in all sub-classes. In Figs. \ref{fig:MBH_10kpc_ correlns_full_sample}, \ref{fig:MBH_3Re_ correlns_full_sample} and \ref{fig:MBH_5Re_ correlns_full_sample}, we see very similar results 
also for the central 10 kpc, 3$R_{\rm e}$ and 5$R_{\rm e}$ regions, respectively. While for the BCGs the $M_{\rm BH}$-$kT$ correlation is found to be very strong at all radii, for the 
non-BCG ellipticals and lenticulars, the same correlation is found to be weak at $R_{\rm e}$ and becomes even worse as we move out to larger radii. 
For the BCGs, $M_{\rm BH}$ correlates better with $kT$ than with 
$\sigma_{\rm e}$, at all radii. Note that, the observed $M_{\rm BH}$-$kT$ correlation for the BCGs is also much tighter (scatter $ \lesssim$0.28 dex at all radii) than that with the  
large scale intracluster medium temperature \citep[scatter $\sim$0.4-0.9 dex; see][]{Bogdan2018,Phipps2019}.

\section{Discussion}
\label{sec:discussion}
The well known $M_{\rm BH}$-$\sigma_{\rm e}$ correlation reflects the connection between $M_{\rm BH}$ and the stellar mass in the very centre of the galaxy, where the contribution 
of dark matter to the total mass is negligible. The break-down of the $M_{\rm BH}$-$\sigma_{\rm e}$ correlation at the high mass end, in BCGs seen in the top right panel of 
Fig.~\ref{fig:MBH_Re_ correlns_full_sample}, has been previously discussed \citep{KormendyHo2013,Hilz2012} and attributed 
to post-merger violent relaxation in systems undergoing a large number of mergers. 

\begin{figure*}
\centering
  \includegraphics[width=0.48\linewidth]{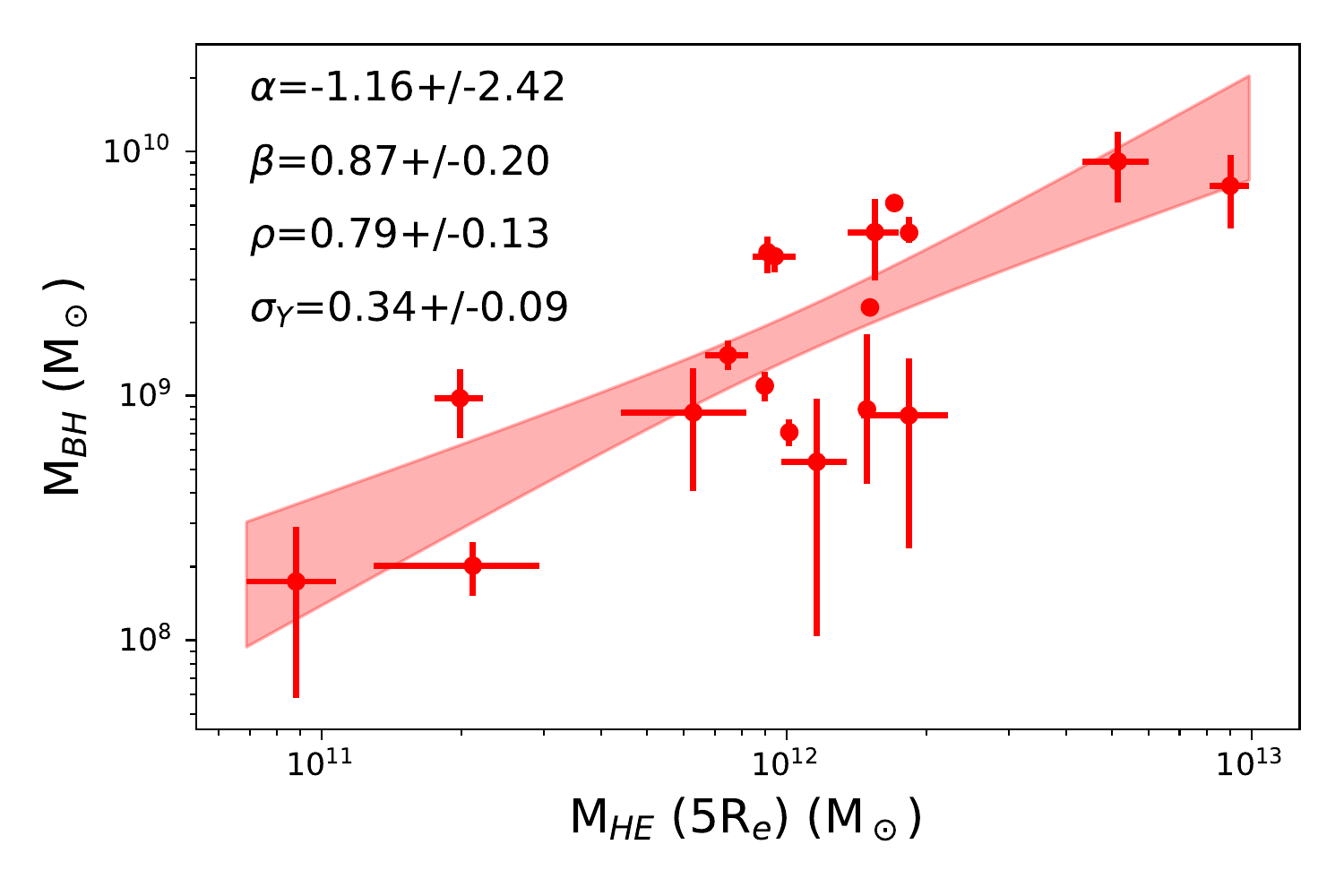}
  \includegraphics[width=0.495\linewidth]{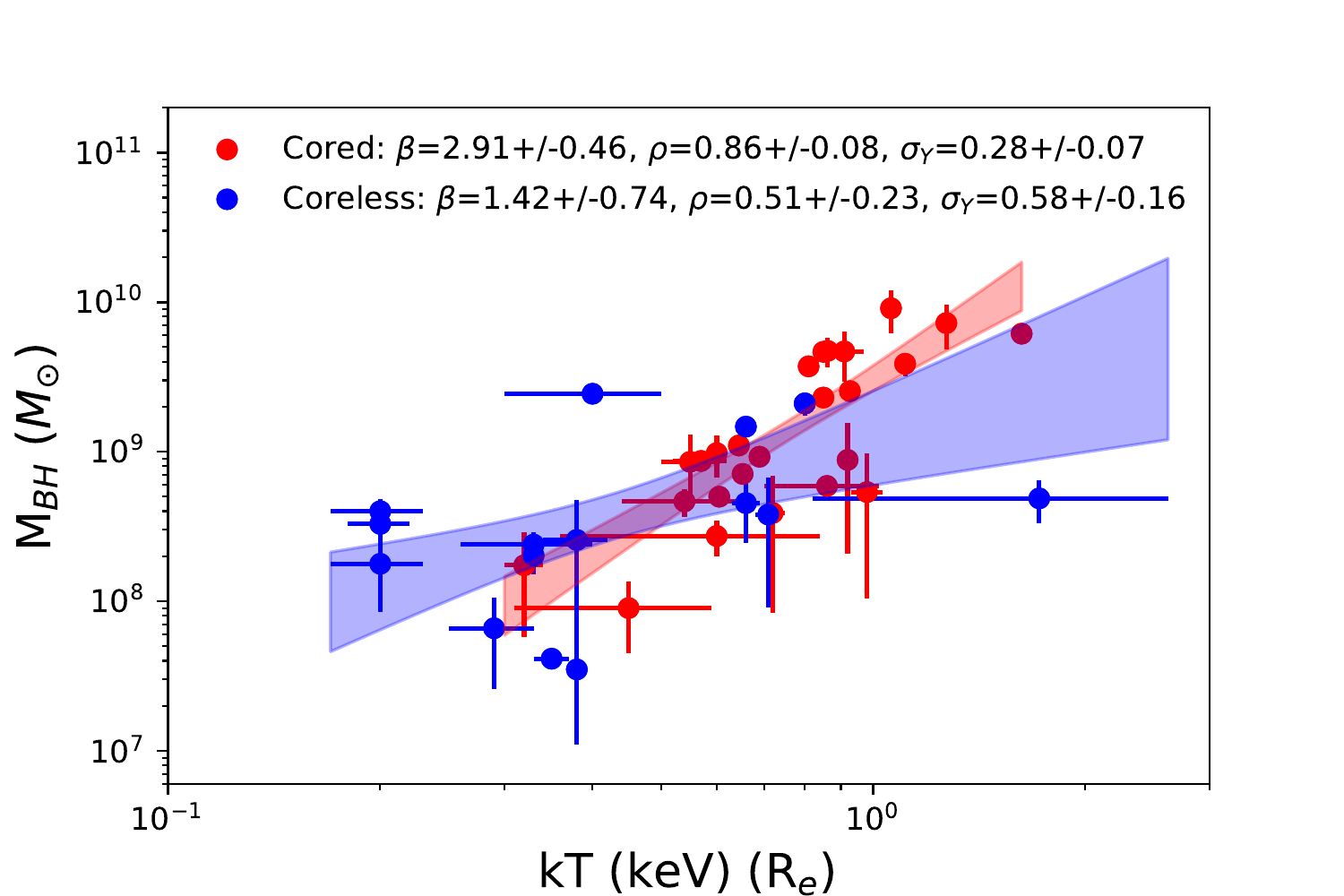}
 \caption{Left: Correlation of the central SMBH mass ($M_{\rm BH}$) with the hydrostatic mass ($M_{\rm HE}$) estimated within 5$R_{\rm e}$ for the BCGs. The shaded area shows the 68\% confidence region for the correlation. Right: Correlations of the SMBH mass with the hot gas temperature (within $R_{\rm e}$) of the 'cored' (red points) and 'coreless' (blue points) galaxies. 
The parameters obtained from the correlation analyses are defined in the same way as in Fig. \ref{fig:MBH_Re_ correlns_full_sample}, and their 68\% uncertainties are given in the insets in both of the panels.}
\label{fig:BCGs_MBH_Mtot_correln}
\end{figure*}

In hydrostatic equlibrium, $kT$ reflects the total mass ($M_{\rm tot}$) of the galaxy and therefore, next to the strong $M_{\rm BH}-kT$ correlation, we also expect a 
correlation between $M_{\rm BH}$ and $M_{\rm tot}$ for BCGs. We tested this by measuring the hydrostatic masses ($M_{\rm HE}$) of the BCGs 
within 5$R_{\rm e}$, calculated using the relation $M_{\rm HE} (<R) = -(R^2/G \rho_{\rm g})\;dP/dr$ at $R=5R_{\rm e}$; where $G$ is the gravitational constant.
The gas mass density ($\rho_{\rm g}$) and the 
pressure gradient ($dP/dr$) at $R$ were determined by fitting the deprojected mass density and pressure profiles with powerlaw model fits. As shown in 
Fig. \ref{fig:BCGs_MBH_Mtot_correln} (left), we find a moderately strong and almost linear (slope of $0.87\pm0.20$) correlation between $M_{\rm BH}$ and $M_{\rm HE} (<5 R_{\rm e})$, 
with a coefficient of 0.79$\pm$0.13 and a scatter of 0.34$\pm$0.09 dex. We also calculated the hydrostatic masses within $R_{\rm e}$ and 3$R_{\rm e}$ and obtained similar $M_{\rm BH}$-$M_{\rm HE}$ 
correlations (see Fig. \ref{fig:BCGs_MBH_Mtot_Re_3Re_correln}). 
The observed scatter could be the result of the inherent uncertainty in the hydrostatic mass measurements due to non-thermal pressure support, which on average contributes $\sim20$--30\% of the thermal 
pressure \citep{Churazov2010}. However, it may also be contributed by additional effects such as abundance gradients, uncertainties in the estimation of $R_{\rm e}$ and other systematic 
uncertainties \citep{Paggi2017}. At the 
effective radii of BCGs, dark matter starts to become a significant component of the total mass budget \citep[see][]{Barnabe2011,Lovell2018}\footnote{The observational measurements of 
DM fraction ($f_{\rm DM}$) within $R_{\rm e}$ are strongly sensitive to the adopted stellar initial mass function (IMF). For example, \citet{Barnabe2011} show that for the most massive 
early type galaxies ($M_*>10^{10}$ $M_{\odot}$), similar to our BCGs, adopting the Salpeter IMF results in $f_{\rm DM}$ ($<R_{\rm e}$) of less than 50\% while adopting the Chabrier 
IMF results in a significantly higher $f_{\rm DM}$ (>60\% ).} and at larger radii of $3R_{\rm e}$ and $5R_{\rm e}$, it becomes the dominant component. The observed moderately strong 
and almost-linear correlation between  $M_{\rm BH}$ and $M_{\rm tot}$, therefore, indicates that for the most massive galaxies in the centres of groups/clusters, the mass of the central 
SMBH also correlates with the underlying dark matter halo mass ($M_{\rm DM}$). 

We interpret this result in the light of state-of-the-art cosmological simulations. As discussed in the introduction,
using results from the TNG300 simulations, \citet{Weinberger2018} found that for the growth of the most massive SMBHs ($M_{\rm BH}>10^{8.5}$ $M_{\odot}$), mergers are the most important 
channel, with the most massive SMBHs ($M_{\rm BH}>10^{10}$ $M_{\odot}$) undergoing $\sim$1000 mergers, presumably following the mergers of their host galaxies.
Given that the BCGs in our sample are among the most massive galaxies in the Universe, the observed near-linear correlation between the total mass of the galaxy and the central SMBH 
could be a natural consequence of the non-causal coevolution of SMBHs and galaxies through mergers \citep{Weinberger2018}.

To investigate this further, we divided our sample into ``core'' and ``coreless'' galaxies based on the shape of their optical surface-brightness profiles. Using 
the information available in the literature \citep[][]{Pellegrini2005,Lauer2007,Hopkins2009,Krajnovic2013,KimFabbiano2015,Saglia2016,Forbes2017}, we found 25 core and 16 coreless galaxies in our sample (most BCGs and slow-rotating non-BCG ellipticals are core galaxies, while the S0s and other fast rotating ellipticals are typically coreless). 
The $M_{\rm BH}$-$kT$ relations for the core and coreless galaxies are shown in Fig. \ref{fig:BCGs_MBH_Mtot_correln} (right). 
The core galaxies are found to have a very strong $M_{\rm BH}$-$kT$ correlation (correlation coefficient $\sim$0.9, intrinsic scatter $\sim$0.3) while the coreless galaxies 
are found to have a weak $M_{\rm BH}$-$kT$  correlation (correlation coefficient $\sim$0.5, intrinsic scatter $\sim$0.6). The cores seen in the optical surface-brightness profiles 
of mostly elliptical galaxies, are believed to 
have formed by the stellar excavation accompanying the orbital decay of the binary SMBHs that form as a result of dissipationless (``dry") mergers of their host galaxies. The 
coreless galaxies, on the other hand, are believed to have gone through dissipative (``wet") mergers \citep{Kormendy2009}. The observed tight $M_{\rm BH}$-$kT$  
correlation for the core galaxies could therefore be a natural consequence of the large number of dry galaxy mergers followed by the gradual inspiral and merger of their SMBHs.

The $M_{\rm BH}$-$kT$ correlation could also be affected by AGN feedback. 
The $M_{\rm tot}$-$kT$ correlation for early type galaxies is found to have a much steeper slope than 
the value expected based on self-similar evolution \citep[$\sim$1.5; see][]{Babyk2018}. The relativistic jets emanating from the central AGN in BCGs are likely to increase the temperature with 
respect to the virial equilibrium value. This effect is expected to be more pronounced in the low-mass galaxies than the high-mass systems \citep{Goulding2016}, which can explain the 
steepening of the $M_{\rm tot}$-$kT$ relation \citep{Babyk2018}. While it is non-trivial to assess 
the exact impact of AGN feedback on the host galaxies, it most likely plays a secondary role in producing the observed $M_{\rm BH}$-$kT$ correlation.

The low-mass non-BCGs and lenticular galaxies do not show a strong $M_{\rm BH}$-$kT$ correlation. This may imply that because of the fewer mergers, the applicability of 
the central limit theorem is limited and the $M_{\rm BH}$-$M_{\rm tot}$ correlation is weaker in these low-mass systems. However, the lack of the $M_{\rm BH}$-$kT$ relation does not need 
to necessarily imply the lack of the $M_{\rm BH}$-$M_{\rm tot}$ correlation.  In fact, the modest $M_{\rm BH}$-$L_{\rm X}$ correlation observed for all three classes of galaxies 
suggests that the $M_{\rm BH}$-$M_{\rm tot}$ correlation might be in place for all galaxies in our sample. Due to their smaller gravitational potentials, non-BCGs and lenticulars 
are more sensitive to feedback \citep{KimFabbiano2013,KimFabbiano2015,Goulding2016} and for a significant fraction of the lower mass galaxies, the X-ray atmospheres may be in an outflow state \citep{Pellegrini2018}. Ram-pressure stripping of 
non-BCG galaxies moving through the intra-cluster medium can remove a significant fraction of the galactic atmospheres, altering the thermal state of the gas. These systems, 
therefore, may not be in hydrostatic equilibrium and their atmospheric temperatures may not accurately reflect the gravitational potential of the host galaxies. 

Finally, although mergers and central limit theorem can provide a plausible non-causal explanation for the observed $M_{\rm BH}$-$kT$/$M_{\rm BH}$-$M_{\rm tot}$ correlation, 
a pure merger picture has a difficulty to account for a class of massive compact `relic' galaxies. These galaxies observed at z$\sim$0, are likely the remnants of the 
progenitors of giant ellipticals, which remained largely untouched by mergers since their early, rapid dissipative growth \citep[e.g.][]{Trujillo2014,FerreMateu2017,Werner2018,Buote2018,Buote2019}. 
Fig. \ref{fig:MBH_Re_ correlns_full_sample} shows that the two 
well-studied massive `relic' galaxies, NGC 1277 and Mrk 1216, lie on the tight $M_{\rm BH}$-$kT$ correlation for BCGs. This result is puzzling, because those relic 
galaxies are believed to have avoided the stage of growth by dry mergers, which suggests that an initial correlation between SMBHs and 
DM halos could have already existed early in the evolution of these galaxies, following the formation of their X-ray atmospheres. This is consistent with the predictions of the EAGLE simulations, where the rapid growth of black holes following the formation of hot atmospheres \citep[see][]{Bower2017,McAlpine2018} results in an $M_{\rm BH}$ which is determined by the binding energy of the halo \citep[see][]{booth2010,booth2011}. For BCGs, such an initial correlation could have been subsequently strengthened 
via numerous gas-free mergers with galaxies hosting central SMBHs. For high-mass galaxies, this scenario implies an $M_{\rm BH}$-$M_{\rm tot}$ 
correlation throughout most of the lifetime of the Universe. 

\section{ACKNOWLEDGEMENT}

We are grateful to the referee for the constructive feedback that helped us improve the manuscript significantly. 
This work was supported by the Lend{\"u}let LP2016-11 grant awarded by the Hungarian Academy of Sciences. The scientific results reported here are 
based to a significant degree on data, software and web tools obtained from the High Energy Astrophysics Science Archive Research Center (HEASARC), a service 
of the Astrophysics Science Division at NASA/GSFC and of the Smithsonian Astrophysical Observatory's High Energy Astrophysics Division. The results 
are based on observations made by the \textit{Chandra} X-ray Observatory. We thank J. Kormendy, E. Churazov and F. Mernier for 
useful discussions.

{\footnotesize\bibliography{ref}}

\appendix{}

\setcounter{secnumdepth}{-1}
\renewcommand{\thesection}{A}
\renewcommand\thefigure{\thesection.\arabic{figure}}

\section{}

\begin{figure*}
\centering
  \includegraphics[width=0.7\linewidth]{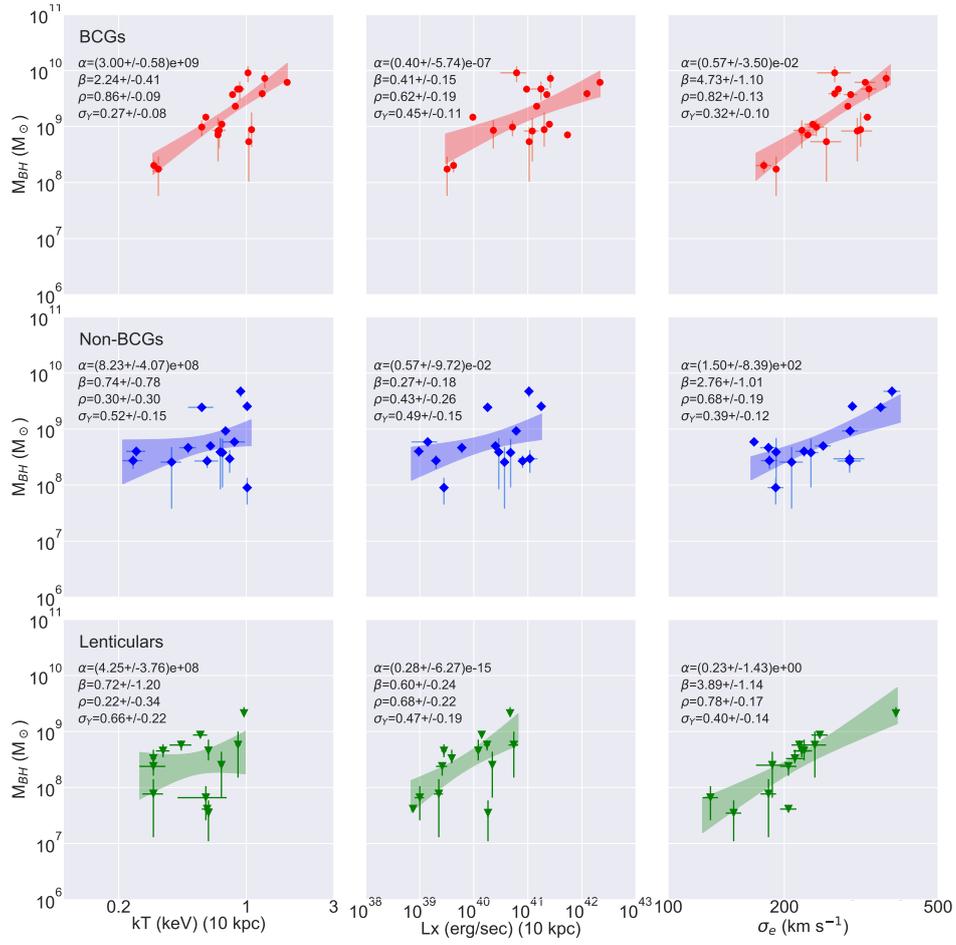}
\caption{Same as Fig. \ref{fig:MBH_Re_ correlns_full_sample}, but for temperatures and luminosities obtained within $r=10$ kpc.}
\label{fig:MBH_10kpc_ correlns_full_sample}
\end{figure*}

\begin{figure*}
\centering
  \includegraphics[width=0.7\linewidth]{Fig4.pdf}
\caption{Same as Fig. \ref{fig:MBH_Re_ correlns_full_sample}, but for temperatures and luminosities obtained within 3$R_{\rm e}$. }
\label{fig:MBH_3Re_ correlns_full_sample}
\end{figure*}

\begin{figure*}
\centering
  \includegraphics[width=0.7\linewidth]{Fig5.pdf}
\caption{Same as Fig. \ref{fig:MBH_Re_ correlns_full_sample}, but for temperatures and luminosities obtained within 5$R_{\rm e}$.}
\label{fig:MBH_5Re_ correlns_full_sample}
\end{figure*}

\begin{figure*}
\centering
  \includegraphics[width=0.48\linewidth]{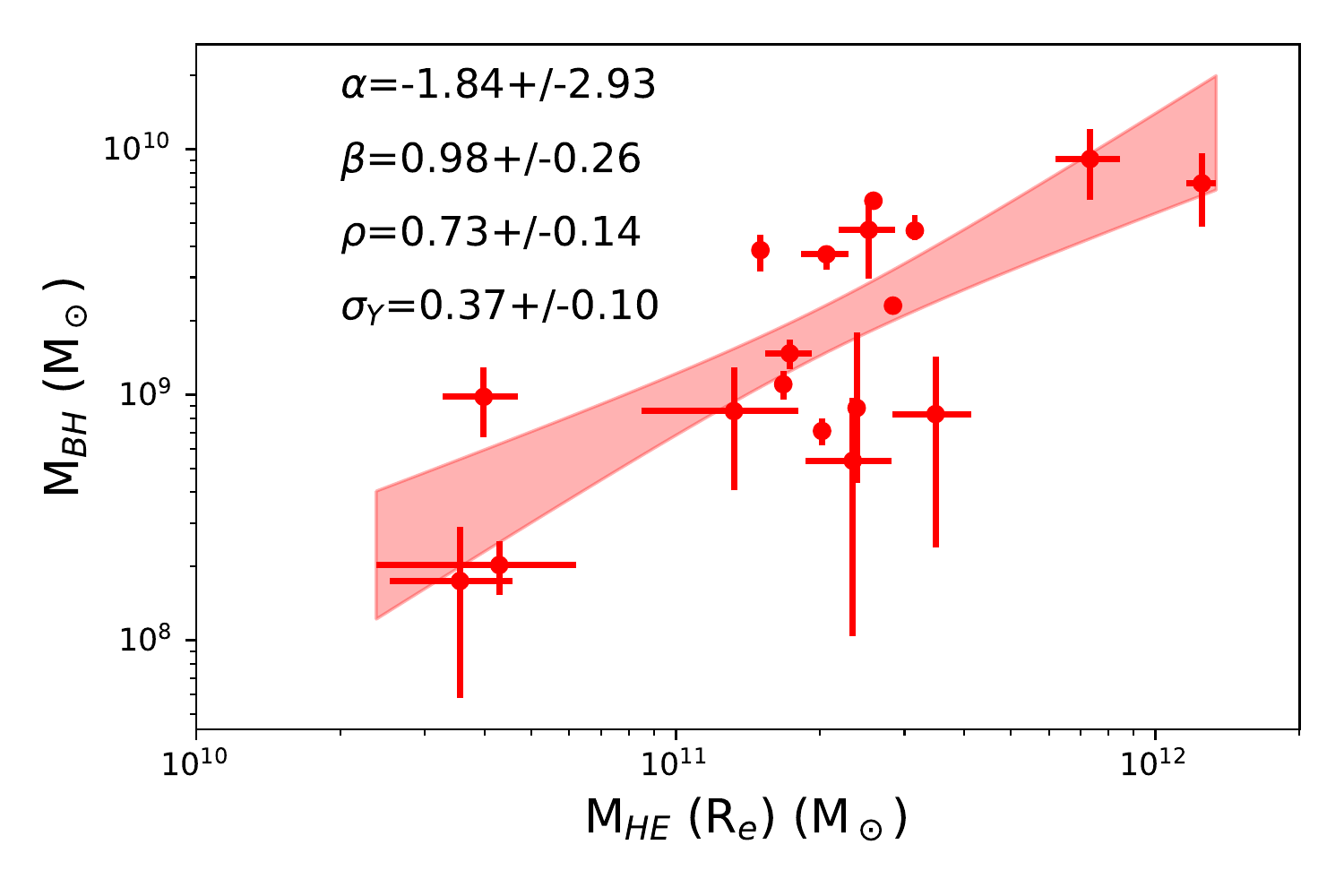}
  \includegraphics[width=0.495\linewidth]{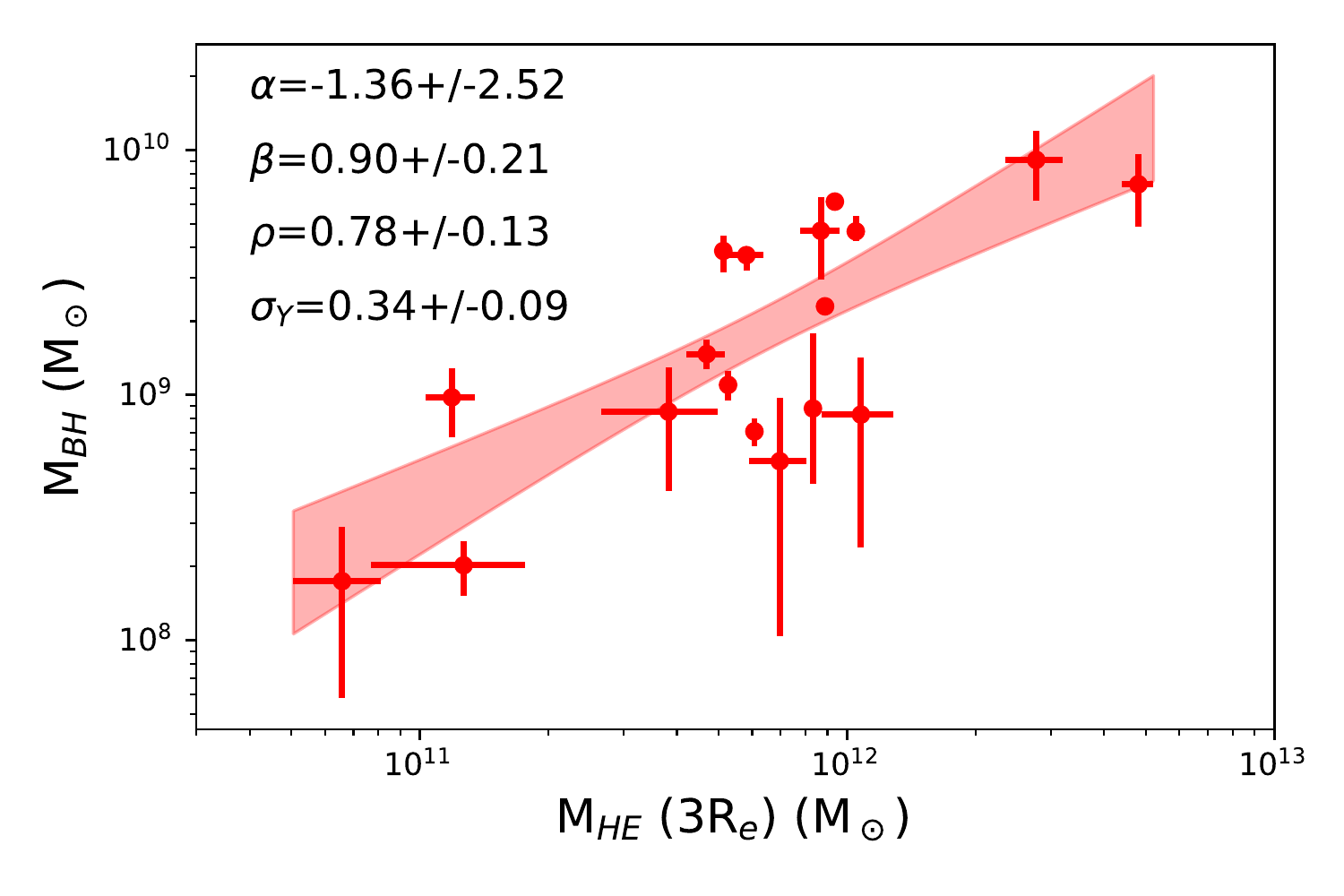}
 \caption{Same as Fig. \ref{fig:BCGs_MBH_Mtot_correln} (left) but with the hydrostatic masses obtained within $R_{\rm e}$ (left) and 3$R_{\rm e}$ (right).}
\label{fig:BCGs_MBH_Mtot_Re_3Re_correln}
\end{figure*}

\end{document}